\documentclass[aps,prl,showpacs,twocolumn,superscriptaddress]{revtex4-2}%

\usepackage{graphicx}
\usepackage{dcolumn}
\usepackage{bm}

\usepackage{float}  

\usepackage{graphicx}
\usepackage{graphics}
\usepackage[caption=false]{subfig}
\usepackage{color}
\usepackage[english]{babel}
\usepackage[utf8]{inputenc}
\usepackage{ae}
\usepackage{latexsym}
\usepackage{dcolumn}
\usepackage{bm}
\usepackage[table]{xcolor}
\usepackage{array}
\usepackage{url}
\usepackage{epsf}
\usepackage{epsfig}
\usepackage{pstricks,pst-grad}
\usepackage[pdftex,colorlinks]{hyperref}
\usepackage{amsmath}
\usepackage{amssymb}
\usepackage{ragged2e}
\usepackage{comment}

\begin{document}

\title{Ion Correlation-Driven Hysteretic Adhesion and Repulsion between Opposing Polyelectrolyte Brushes}

\author{Chao Duan}
\affiliation{Department of Chemical and Biomolecular Engineering, University of California Berkeley, CA 94720, USA}

\author{Rui Wang}
\email {ruiwang325@berkeley.edu}\affiliation{Department of Chemical and Biomolecular Engineering, University of California Berkeley, CA 94720, USA}
\affiliation{Materials Sciences Division, Lawrence Berkeley National Lab, Berkeley, CA 94720, USA}

\date{\today}

\begin{abstract}
Polyelectrolyte (PE) brushes are widely used in biomaterials and nanotechnology to regulate surface properties and interactions. Here, we apply the electrostatic correlation augmented self-consistent field theory to investigate the interactions between opposing PE brushes in a mixture of 1:1 and 3:1 salt solutions. Our theory predicts hysteretic feature of the normal stress induced by strong ion correlations. In the presence of trivalent ions, the force profile is discontinuous: repulsive in the compression branch and adhesive in the separation branch. The molecular origin of the hysteretic force is the coexistence of two collapsed modes: two separated condensed layers on each surface in the compression and a single bundled condensed layer in the separation. With the systematic inclusion of ion correlations, our theory fully captures the hysteretic force, adhesive separation,
``jump-in'' and ``jump-out'' features, and the ``specific ion effect'', all in good agreement with the reported experimental results.



\end{abstract}

\maketitle

\section{1. Introduction}
The applications of polyelectrolyte (PE) brushes are increasingly emerged and expanded to biomaterials and nanotechnology. \cite{Ayres_2010,Chen_2017} The heart of the research is to study the forces between opposing PE brushes, which regulate a wealth of structure and dynamic properties, such as lubrication of articular cartilages, stability of colloidal suspensions, and interaction of protein-coated surfaces. \cite{Raviv_2003,Claesson_2005,Chen_2009,Ohta_2016} Charged polymers are sensitive to external stimuli ranging from mechanical and electrical to environmental changes. \cite{R_he,BALLAUFF20071135,LEE201024} It is desired to understand how PE brushes behave when they are manipulated to tune surface forces such that we can fully realize their potential as surface regulators. \cite{Das_2015}

The addition of salt ions is an important tool to modulate the morphology and mechanical properties of PE brushes. \cite{Romet-Lemonne:2004wx,Mei:2006vx}
It is known that the force between two opposing brushes in monovalent salt solutions is non-hysteretic repulsive, well-captured by the mean-field Poisson-Boltzmann theory. \cite{Messina_2009,dean2014electrostatics} However, the behaviors in the presence of multivalent ions are found quite non-trivial. In a series of experiments conducted by Tirrell group, it was reported that the friction force between PE brushes increases dramatically in multivalent salt solutions, diminishing the lubricity.  \cite{Yu_2018} Sharp decrease in the brush height was simultaneously observed.  \cite{Yu_2016,Yu_2017,Jackson_2017} Using surface forces apparatus (SFA), Farina et al. measured the normal forces between two grafted layers of poly(sodium styrenesul-fonate) (PSSNa) in a mixture of monovalent and trivalent salt solutions. \cite{Farina_2013,Farina_2015} For a given total ionic strength, hysteretic feature emerges above a critical concentration of trivalent cations. The separation and compression processes are no longer reversible but exhibit distinct force profiles. While repulsive force was detected during the compression, adhesive force was observed in the separation, in stark contrast to our common knowledge of the interaction between two like-charged objects. More interestingly, the repulsive force in the compression is dramatically reduced upon increasing the fraction of trivalent ions, which can even continuously turn to adhesive (defined as ``jump-in'' in the Ref. \cite{Farina_2015}). On the other hand, the adhesion in the separation process exhibits a ``jump-out'' feature: it becomes stronger in the early stage but suddenly disappears at a critical distance, indicated by a rapid separation of the two surfaces. Furthermore, besides the salt concentration effect, the hysteretic force profiles also show ``specific ion effect'', which depends on the chemical identity of the trivalent cations. \cite{Farina_2015}
As La$^{3+}$ is replaced by larger ${\rm Ru(NH_3)_6}^{3+}$ or hydrated Al$^{3+}$, the ``jump-in'' signal in the compression process disappears, while the ``jump-out'' in the separation process is postponed to a larger critical distance. All these phenomena cannot even be qualitatively explained by mean-field electrostatics.

Evidences indicate that electrostatic correlation, an obvious factor missing in the mean-field theories, plays a key role in systems containing multivalent ions. Many efforts have been made to include ion correlations in modeling PEs. Muthukumar developed a variational method based on a two-state model for counterion binding. \cite{Muthukumar_2004,Kundagrami_2008}
Applying renormalized Gaussian fluctuation theory, Shen and Wang addressed the electrostatic fluctuation coupled with chain connectivity through quantifying the self energy of a PE. \cite{Shen_2017,Shen_2018} Sing and Qin performed a cluster expansion to study the correlation effect on solution thermodynamics. \cite{Sing_2023}
However, all these approaches were developed to model homogeneous systems, whereas their applicability to inhomogeneous PEs has not been verified yet. Sing and de la Cruz developed a hybrid liquid state integral equation and self-consistent field theory to model correlations in inhomogenous polymer blends and block copolymers. \cite{Sing:2013aa,Sing_2014} However, this approach has to preassume local charge neutrality, which fails wherever a double layer with local charge-separation exist, such as PE adsorption and brushes. To specifically describe PE brushes, Kegler et al. applied a phenomenological model to account for the correlation-induced counterion condensation. \cite{Kegler:2008ue,Jusufi_2001,Jusufi2004,Schneider_2010} An ``effective charge'' was introduced as a fitting parameter. Despite of its simplicity, this method fails to capture the hysteretic feature of the force profile, particularly the adhesion in the separation process. Using a classical density functional theory (cDFT), Prusty et al. successfully predicted the adhesive force between opposing PE brushes in the presence of multivalent ions. \cite{Prusty_2024} However, their force profiles are always continuous, inconsistent with the hysteretic feature observed in the experiments of Farina et al.

In our earlier work, we developed a new theory which systematically incorporates the variational Gaussian renormalized fluctuation theory for electrostatics into the self-consistent field theory (SCFT) for polymers. \cite{Duan:2024aa} The theory is particularly capable of modeling inhomogeneous PEs with spatially varying ionic strength or dielectric permittivity. Applied to a single layer of PE brushes, the theory successfully captures the ion correlation-induced non-monotonic salt concentration dependence of brush height and lateral microphase separation. In the current work, we apply the theory to study the interactions between two opposing PE brushes. Our theory fully captures the hysteretic feature of the force: repulsion in the compression and adhesion in the separation, in good agreement with the experimental results reported by Tirrell group.

\section{2. Theory}

We consider a system of two parallel plates located at $x=\pm D/2$, respectively, where $D$ is the separation distance between the two plates. Each plate is uniformly grafted by $n$ PE chains with the grafting density $\sigma$. Each chain is consisting of $N$ monomers with Kuhn length $b$ and volume $v$. Without loss of generality, we consider each monomer carries one negative charge, $z_P=-1$. The PE grafted plates are immersed in an electrolyte solution containing $n$ types of salt. Each salt $i$ ($i$ ranges from 1 to $n$) has cations of valency $z_{i_+}$ and anions of valency $z_{i_-}$. Cations thus play the role of counterions with regard to the PE charge. The electrolyte solution between the two plates is connected to a bulk reservoir of ion concentration $c^b_{i_\pm}$ to maintain the chemical potentials of solvent $\mu_S$ and ions $\mu_{i_\pm}$. To avoid overestimation of ion correlations caused by the point-charge model, we describe the ionic charge by a finite spread function $h_K({\bf r},{\bf r}^{\prime})$ ($K= P, i_\pm$). \cite{Wang:2010wk,Agrawal_2022,Agrawal:2022ux,Agrawal_2023,Agrawal_2024,Duan:2024aa} In principle, the form of $h_K$ can be arbitary as long as the Born radius of the ion $a_{K}$ is retained. Here, we take $h_K$ to be Gaussian for convenience. The smeared charge model for PE widely used in existing theories fails to capture any correlations between charged monomers as well as between them and mobile ions. We adopt a discrete Gaussian
chain model such that each monomer is considered as a charged particle. The hydrophobic interaction between polymer and solvent is described by the Flory-Huggins $\chi$ parameter.

The improvement of our new theory to the existing work is a non-perturbative variational approach which accounts for the electrostatic fluctuation via introducing a general Gaussian reference action. The consequence of this fluctuation is the self energy of both charged monomers and mobile ions. We refer interested readers to our earlier paper for the detailed derivation. \cite{Duan:2024aa} The key result of our theory is a set of self-consistent equations for polymer density $\phi_P(\bf r)$, conjugate fields of polymer $\omega_P(\bf r)$ and solvent $\omega_S(\bf r)$, mean electric potential $\psi(\bf r)$, ion concentration $c_{i_\pm}(\bf r)$, self energy of charged species $u_K(\bf r)$ $(K=P,i_{\pm})$, and electrostatic correlation function $G(\bf r,{\bf r}^{\prime})$:

\begin{subequations}
\begin{align}
\phi_P({\bf r}) &= \sum_{\alpha=1,2} \frac{n}{Q_{\alpha}} \int_0^N ds q_{\alpha}^{\dagger}({\bf r};s) q_{\alpha}({\bf r};s)
\end{align}
\begin{align}
1-\phi_P({\bf r}) &= e^{\mu_S-\omega_S({\bf r})}
\end{align}
\begin{align}
\omega_P({\bf r}) - \omega_S({\bf r}) & = \chi [1-2\phi_P({\bf r})] - 
v\frac{\partial \epsilon({\bf r})}{\partial \phi_P({\bf r})}
[\nabla\psi({\bf r})]^2
\end{align}
\begin{align}\label{PB}
-\nabla\cdot[\epsilon({\bf r})\nabla\psi({\bf r})] &= \frac{z_P}{v}\phi_P({\bf r}) + \sum_{K=i_\pm} z_K c_K({\bf r})
\end{align}
\begin{align}
c_{i_\pm}({\bf r}) &=\lambda_{i_\pm} e^{-z_{i_\pm} \psi({\bf r})-u_{i_\pm}({\bf r})}
\end{align}
\begin{align}\label{uK}
u_{P,i_{\pm}}({\bf r}) &= \frac{z^2_{P,i_{\pm}}}{2} \int d{\bf r^{\prime}} d{\bf r^{\prime\prime}} h_{P,i_{\pm}}({\bf r},{\bf r}^{\prime}) G({\bf r}^{\prime},{\bf r}^{\prime\prime}) h_{P,i_{\pm}}({\bf r^{\prime\prime}},{\bf r})
\end{align}
\begin{align}\label{GreenFunc}
-\nabla_{\bf r}\cdot[\epsilon({\bf r})\nabla_{\bf r} G({\bf r},{\bf r}^{\prime})] + 2I({\bf r}) G({\bf r},{\bf r}^{\prime}) &= \delta({\bf r}-{\bf r}^{\prime})
\end{align}
\end{subequations}
$\epsilon({\bf r})=kT \epsilon_0 \epsilon_r({\bf r}) /e^2$ is the scaled permittivity with $\epsilon_0$ the vacuum permittivity, $e$ the elementary charge and $\epsilon_r({\bf r})$ the local dielectric constant that can be evaluated based on the local composition \cite{Sing_2014,Zhuang_2021}.
$\lambda_{i_\pm}=e^{\mu_{i_\pm}}/v_{i_\pm}$  is the fugacity of salt ions with $v_{i_\pm}$ the ion volume.
$I({\bf r})=(1/2)\sum_{K=P,i_\pm} z^2_K c_K({\bf r})$ is the local ionic strength. $q_{\alpha}({\bf r};s)$ and $q_{\alpha}^{\dagger}({\bf r};s)$ are the forward and backward chain
propagators, respectively, for the chain grafted on Plate $\alpha$ ($\alpha=1,2$). \cite{Fredrickson_Book2006} They are determined by 
\begin{subequations}
\begin{align}\label{q_forward}
& \frac{\partial q_{\alpha}({\bf r};s)}{\partial s} =
\frac{b^2}{6} \nabla^2 q_{\alpha}({\bf r};s) - U_P({\bf r}) q_{\alpha}({\bf r};s)
\end{align}
\begin{align}\label{q_backward}
& - \frac{\partial q_{\alpha}^{\dagger}({\bf r};s)}{\partial s} = \frac{b^2}{6} \nabla^2 q_{\alpha}^{\dagger}({\bf r};s) - U_P({\bf r}) q_{\alpha}^{\dagger}({\bf r};s)
\end{align}
\end{subequations}
where $U_P({\bf r})=\omega_P({\bf r})+z_P\psi({\bf r})+u_P({\bf r})$ is the total interaction field experienced by monomers.

\begin{table*}[t]
  \caption{Parameters used to model PSSNa brushes in various salt solutions}
  \label{Parameter}
  \begin{tabular}{ccccccccccc}
    \hline
    $\chi$ & $N$ & $b$ \cite{Yoshizaki_1988,Hirose_1999} & $v$ &  $\sigma$ \cite{Farina_2013,Farina_2015} & ${\epsilon}_r$ & $a_{\rm Na^+}$ \cite{dos_Santos_2010} & $a_{\rm NO^-_3}$ \cite{dos_Santos_2010} & $a_{\rm SO^-_3}$ & $a_{\rm La^{3+}}$ \cite{Farina_2015} & $a_{\rm Al^{3+}}$ \cite{Farina_2015} \\
    \hline
    0 & 140 & 0.70nm & 0.91nm$^3$ & 0.0176/nm$^2$ & 80 & 2.5$\mathring{\rm A}$ & 2.0$\mathring{\rm A}$ & 2.5$\mathring{\rm A}$ & 3.1$\mathring{\rm A}$ & 4.8$\mathring{\rm A}$ \\
    \hline
  \end{tabular}
\end{table*}

The resulting free energy of the system is given by
\begin{align}\label{FreeE}
W& = -\sum_{\alpha=1,2}n{\rm ln}Q_{\alpha}-e^{\mu_S} Q_S \nonumber\\
&+ \frac{1}{v} \int d{\bf r} \left [ \chi \phi_P (1-\phi_P) -\omega_P \phi_P-\omega_S (1-\phi_P)  \right] \nonumber\\
&+\int d{\bf r} \left [ \frac{1}{2}\psi\nabla\cdot(\epsilon\nabla\psi)-c_+ -c_- \right ] \nonumber\\
&+ \int d{\bf r} \sum_{K=P, i_\pm} c_K({\bf r}) \left[ \int^1_0 d{\tau} u_K({\bf r};\tau)-u_K({\bf r}) \right]
\end{align}
where $Q_{\alpha}= v^{-1} \int d{\bf r} q_{\alpha}^{\dagger}({\bf r};s) q_{\alpha}({\bf r};s) $ is the single-chain partition function of PE grafted on Plate $\alpha$. $Q_S=  v^{-1} \int d{\bf r} e^{-\omega_S({\bf r})}$ is the partition funciton of solvent molecules. The last term in Eq. \ref{FreeE} is the electrostatic correlation energy derived from the ``charging method'' with $\tau$ ($0 \le \tau \le 1$) the ``charging parameter'' . \cite{Wang_2015,Agrawal:2022ux}
The intermediate self energy $u_K({\bf r};\tau)$ during the ``charging process'' is calculated from the corresponding Green function $G({\bf r},{\bf r}^{\prime};\tau)$, which is solved by Eq. \ref{GreenFunc} with $I({\bf r})$ replaced by $\tau I({\bf r})$. The normal stress $P$ between the opposing brushes can thus be determined by
\begin{align}\label{P}
P& = -\frac{\partial (W_{exc}/A)}{\partial D}
\end{align}
with $W_{exc}=W - W_b$ the free energy in excess of the bulk salt solution and $A$ the surface area of each plate.

It should be noted that we neglect the contribution of intra-chain correlation to self-energy due to its less importance than the individual contribution of a charged monomer $u_K(\bf r)$ under the condition of high ionic strengths (either high salt concentration or high PE density). \cite{Shen_2017,Duan:2024aa} For simplicity, we only consider the excluded volumes of polymer and solvents, whereas those of mobile ions are neglected.

\begin{figure}[t]
\centering
\includegraphics[width=0.48\textwidth]{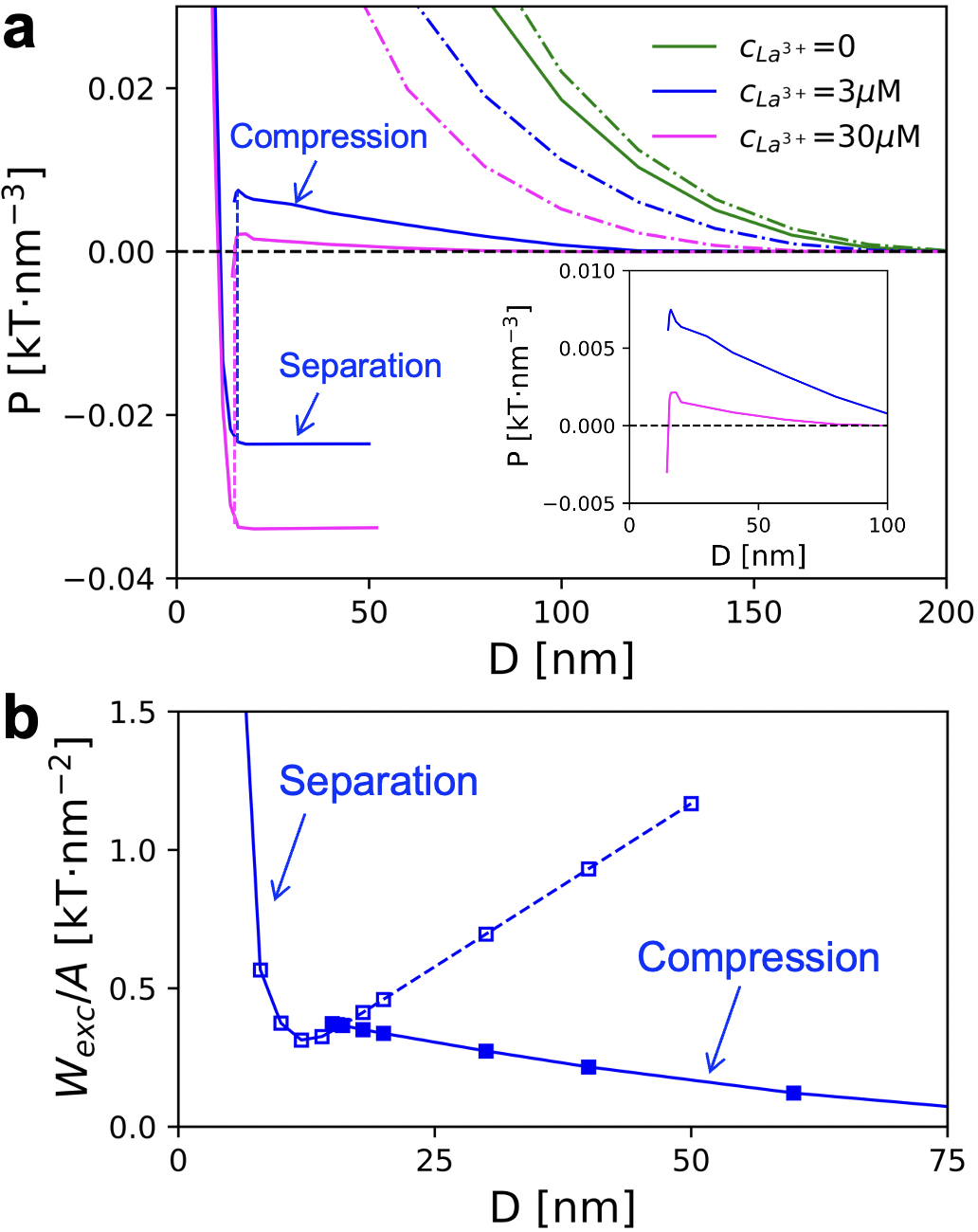}
\caption{
Hysteretic force between two opposing PSSNa brushes in a mixture of monovalent NaNO$_3$ and trivalent La(NO$_3$)$_3$ solutions. (a) Normal stress $P$ as a function of separation distance $D$ for various trivalent salt concentrations. The bulk ionic strength is fixed at $I_b=3$mM. Mean-field results are presented by dashdotted lines for comparison. The vertical dashed lines locates the two coexistent states belonging to the compression branch and the separation branch, respectively. The inset highlights the emergence of ``jump-in'' in the compression process at $c_{{\rm La}^{3+}}=30\mu$M.
(b) The excess free energy per area $W_{exc}/A$  for the compression (filled symbols) and separation (open symbols) at $c_{{\rm La}^{3+}}=3\mu$M. Dashed lines illustrate their corresponding metastable regions.}
\label{fig1}
\end{figure}

\section{3. Results and Discussion}

To elucidate the hysteretic adhesion and repulsion observed by Farina et al., we applied our theory to their experimental system to facilitate a direct comparison. We consider oppsoing poly(sodium styrenesulfonate) (PSSNa) brushes in the mixture of monovalent NaNO$_3$ and trivalent La(NO$_3$)$_3$ (or Al(NO$_3$)$_3$) solutions. To examine the salt concentration effect, we vary the fraction of trivalent salt in the mixture while maintaining the total bulk ionic strength, the same as the experimental setup. Different trivalent cations are also investigated to examine the specific ion effect. The model parameters are summarized in Table \ref{Parameter}.

\begin{figure*}[t]
\includegraphics[width=0.98\textwidth]{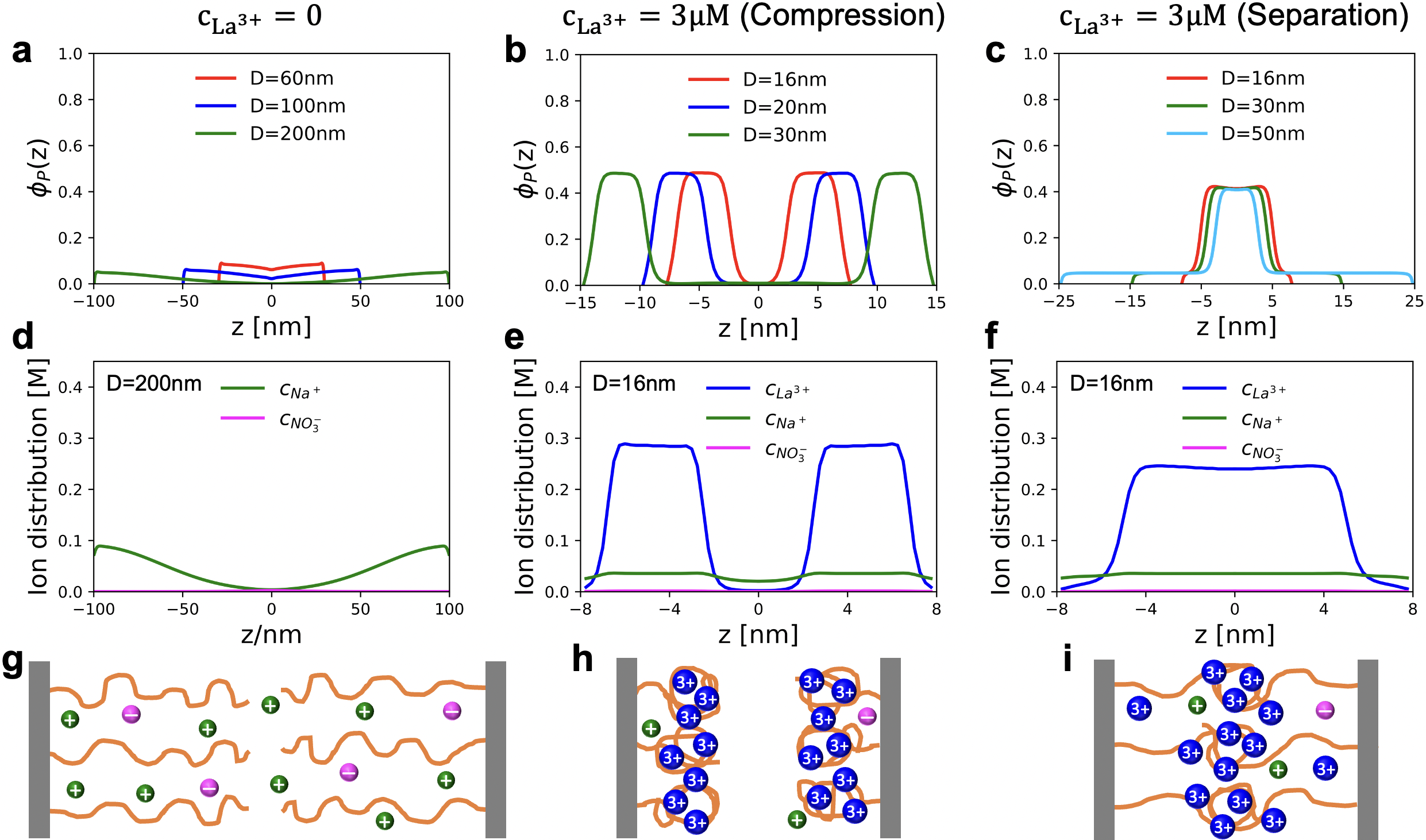}
\caption{
Molecular origin of the hysteretic repulsion and adhesion. Polymer density profiles with different $D$ are plotted for (a) the absence of La$^{3+}$ and $c_{{\rm La}^{3+}}=3\mu$M in the compression (b) and separation (c), respectively. (d), (e) and (f) plot the corresponding ion distributions. (g)-(i) illustrate the corresponding schematics of the brush morphology: the swollen brushes for $c_{{\rm La}^{3+}}=0$, two separated collapsed layers for $c_{{\rm La}^{3+}}=3\mu$M (compression) and a single bundled collapsed layer for $c_{{\rm La}^{3+}}=3\mu$M (separation).
}
\label{fig2}
\end{figure*}

The morphology and free energy of the opposing PE brushes are tracked with respect to the separation distance $D$. Figure \ref{fig1}a plots profiles of normal stress $P(D)$ in a mixture of 1:1 and 3:1 salt solutions with fixed bulk ionic strength $I_b=3$mM. The results based on mean-field electrostatics are also provided for comparison, which predict non-hysteretic repulsion ($P(D)>0$) for all cases. In stark contrast, ion correlations has dramatic impacts on the brush interaction in the presence of trivalent ions. With the inclusion of ion correlations, the normal stress predicted by our theory remains repulsive for pure monovalent salt solution ($c_{{\rm La}^{3+}}=0$). However, when $c_{{\rm La}^{3+}}=3\mu$M, a hysteretic force profile emerges. The entire $P(D)$ becomes discontinous and can be divided into two branches; the compression and separation are no longer on a reversible path. During the compression process, the force is repulsive, but much weaker than the corresponding mean-field result. More intriguingly, the force in the separation process can turn to adhesive, i.e. $P(D)<0$. Furthermore, ion correlations are enhanced as $c_{{\rm La}^{3+}}$ increases to $30\mu$M, which leads to stronger attraction between charged monomers. As a result, the repulsion in the compression is weakened whereas the adhesion in the separation is strengthened. It is interesting to note that $P(D)$ changes sign near the end of the compression branch (as shown in the inset of Figure \ref{fig1}a). The repulsive force thus continuously turn to adhesive, showing a signal of ``jump-in'' as defined in the experiments. Our theory fully captures the hysteretic force, adhesive separation and ``jump-in'' feature, which are all beyond the mean-field prediction. Using the same parameters as the experimental setup, our results are in good agreement with the SFA measurements conducted by Farina et al. \cite{Farina_2013,Farina_2015}

To uncover the nature of the hysteretic repulsion and adhesion, the free energies of the compression and separation branches for $c_{{\rm La}^{3+}}=3\mu$M are plotted in Figure \ref{fig1}b. It can be clearly seen that the free energies of the two branches cross with each other at a $D^*=16$nm, corresponding to the discontinuity in the force profile (see the vertical dash line in Figure \ref{fig1}a). The crossing of the free energy also demonstrates the coexistence of two states at this critical separation distance. According to the feature of discontinuous transition, these two states can extend beyond the crossing point, forming two metastable regions, respectively. Due to the metastability, compression and separation do not need to exactly follow the minimum free energy path. Instead, the real compression and separation (i.e., the forward and backward processes) can undergo different pathways if each includes different portions of the metastable regions. This reveals the origin of the hysteretic force observed in experiments. \cite{Farina_2013,Farina_2015}
Furthermore, it is worth noting that the free energy of the metastable region in the separation branch keep increasing as $D$ increases and becomes much higher than the energy of the compression branch. The system thus gets increasingly unstable as the separation goes on, and tends to undergo a cascade transition to the compression branch. This explains the ``jump-out'' feature of the separation process observed by SFA measurements \cite{Farina_2015}: the adhesion suddenly disappears at a certain $D$, followed by a rapid separation of the two surfaces.

Our theory provides the detailed distribution of both polymers and ions (see Figure \ref{fig2}), enabling us to understand the molecular insight underlying the non-trivial brush interactions. In the absence of trivalent cations, the two opposing PE brushes are highly swollen and repel each other due to the electrostatic repulsion between charged monomers. As shown in Figure \ref{fig2}a, brushes continuously interpenetrate each other as $D$ decreases, leading to the continuous and non-hysteretic repulsive force. Monovalent ions thus can only play the role of charge screening.

Surprisingly, adding a small amount of trivalent cations causes a dramatic change of the brush morphology. While the strong condensation of trivalent ions driven by ion correlations induces brushes to collapse, the exact mode of brush collapse depends on the operation process: compression or separation. In the compression process, the collapse occurs independently in each of the two brushes, as illustrated by the schematic in Figure \ref{fig2}h. There is almost absence of interference and interpenetration of the two collapsed layers as they approach each other, demonstrated by the nearly unchanged polymer density profile of the collapsed layer in Figure \ref{fig2}b. This largely weakens the repulsive force during the compression. On the other hand, in the separation process, a single bundled layer of polymers is formed in the middle of the channel, driven by the emergence of a trivalent ion bridging (see Figure \ref{fig2}i). Two polymer brushes strongly interpenetrate each other, which is the origin of the adhesive force in the separation. When the two plates are separated, the bundled layer gradually melts to swollen brushes near the plate surface as shown in Figure \ref{fig2}c. The middle bundled layer becomes increasingly thinner as more polymers are melted, triggering the occurrence of ``jump-out''.

\begin{figure}[t]
\centering
\includegraphics[width=0.48\textwidth]{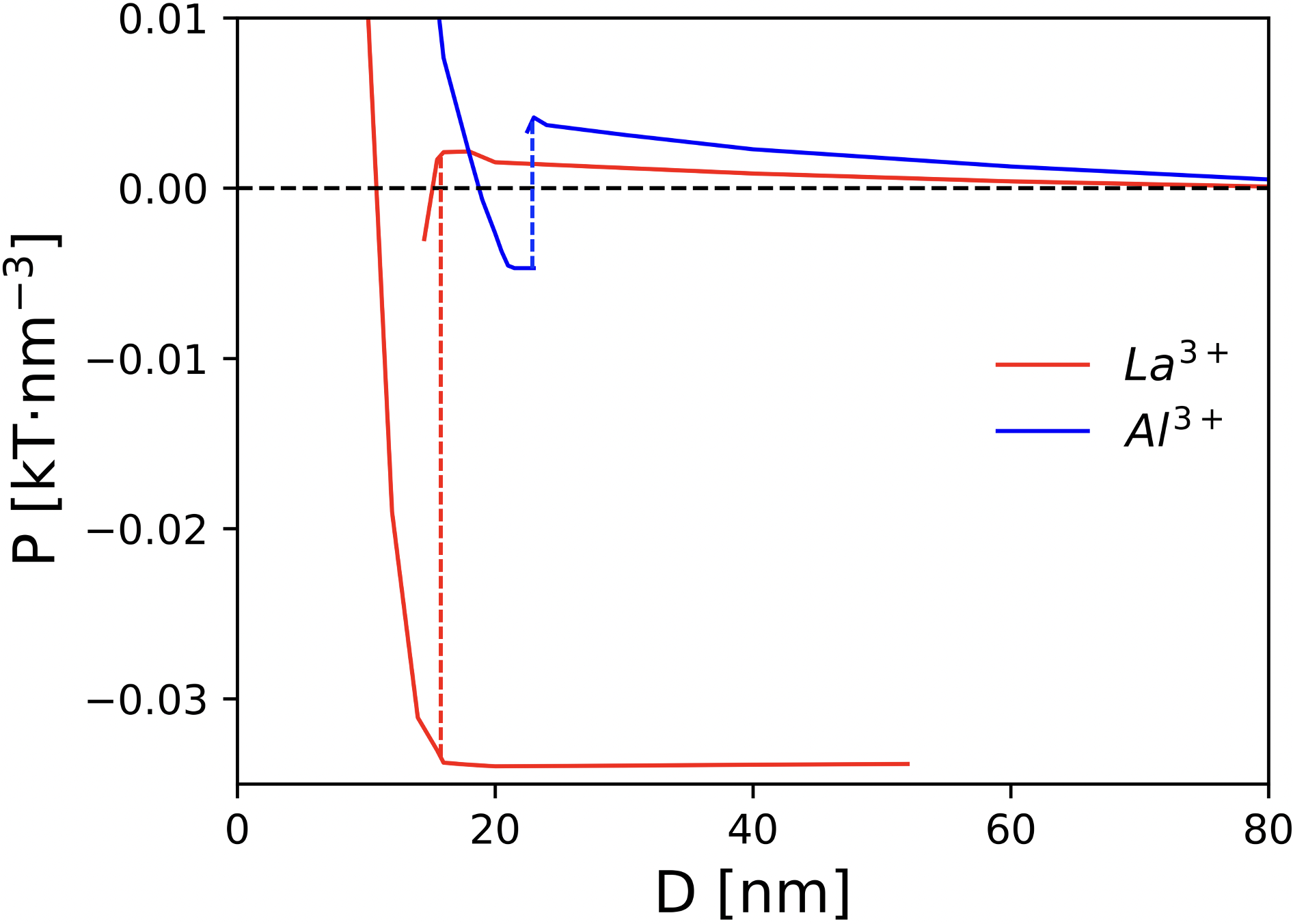}
\caption{
The specific trivalent ion effect on the profiles of normal stress between opposing PSSNa brushes in a mixture of NaNO$_3$ and La(NO$_3$)$_3$ (or Al(NO$_3$)$_3$) salt solutions. $c_{3+}=30\mu$M and the total bulk ionic strength is maintained at $I_b=3$mM.
}
\label{fig3}
\end{figure}

Our results clearly elucidate that the molecular origin of the hysteretic force is the coexistence of two collapsed modes induced by strong ion correlations. When compressing two brushes from large $D$, two condensed layers are formed separately on each plate. They repel each other without noticing the existence of another possible mode of a single bundled layer. On the other hand, when separating a joint condensed layer from small $D$, the adhesive force between two highly interpenetrating brushes can maintain the mode of a single bundled layer for a certain range of $D$ without breaking. It is also worth noting that the condensation of PE brushes are triggered by ion correlations. Even a small amount of trivalent ions (e.g. 0.1$\%$ with respect to the total ionic strength) can completely turns the major role of salts from charge screening to counterion bridging. The latter effect, totally missing in the mean-field electrostatics, is fully captured by our theory.

While the electrostatic correlation is the key factor for the hysteretic force, its strength depends on the chemical identity of ions, usually manifested as the ``specific ion effect''. Experiments of Farina et al. found that as La$^{3+}$ is replaced by larger (Ru$(\rm NH_3)_6$$^{3+}$) or hydrated Al$^{3+}$, the ``jump-in'' signal in the compression process disappears and the ``jump-out'' in the separation process is postponed to a larger critical distance. \cite{Farina_2015} This can be explained by the fact that ions with larger radius for the charge spread have weaker Coulomb interactions and thus less significance in ion correlations. Figure \ref{fig3} plots the normal stress profiles in the presence of La$^{3+}$ and Al$^{3+}$, respectively. Our results show that, the repulsion in the compression process for the case of Al$^{3+}$ is stronger than La$^{3+}$. The entire compression branch is repulsive and the ``jump-in'' signal disappears. Meanwhile, the adhesive force in the separation branch gets weaker. The location of the coexistence shifts to a larger separation distance, which may imply the postpone of the ``jump-out'' signal in the separation process observed in the experiments. The smaller ion correlations of larger trivalent ions reduces the formation of ion bridging, thus leading to both the increase of repulsion and decrease of adhesion. Furthermore, the smaller ion correlations also make the hysteretic feature of the force less pronounced. This is indicated by the smaller gap of the discontinuity and shortened metastable regions as shown in Figure \ref{fig3}. It demonstrates again that the hysteretic adhesion and repulsion is of the electrostatic origin, driven by the strong correlations of multivalent ions.

\section{4. Conclusions}

In this work, we apply the electrostatic correlation augmented self-consistent field theory to investigate the interactions between opposing PSSNa brushes in the mixture of monovalent NaNO$_3$ and trivalent La(NO$_3$)$_3$ solutions. Our theory predicts hysteretic feature of the normal stress when a small amount of trivalent ion (e.g. 0.1$\%$ with respect to the total ionic strength) is added. The force profile is discontinuous: repulsive in the compression branch and adhesive in the separation branch. We also elucidate that the molecular origin of the hysteretic is the coexistence of two collapsed modes induced by strong ion correlations. In the compression process, two condensed layers are formed separately on each plate without interpenetration. In the separation process, a single bundled layer driven by trivalent ion bridging is formed in the middle of the channel, where brushes strongly interpenetrate with each other. 
The melting of the bundled layer leads to the emergence of ``jump-out'' in the separation process. With the increase of La$^{3+}$ concentration, the repulsive force in the compression continuously turns to adhesive, showing a signal of ``jump-in''. Furthermore, we shows the ``specific ion effect'' of the hysteretic force.
As La$^{3+}$ is replaced by larger hydrated Al$^{3+}$, ion correlations become less significant, leading to stronger repulsion in the compression and weaker adhesion in the separation. The ``jump-in'' signal in the compression disappears and the ``jump-out'' in the separation is postponed to a larger critical distance.

We demonstrates the key role of ion correlations on the hysteretic adhesion and repulsion. This effect is totally missing in the mean-field electrostatics but fully captured by our theory. The hysteretic force, adhesive separation, ``jump-in'' and ``jump-out'' features predicted by our theory are in good agreement with the experimental results reported by Tirrell group. Our theory can be easily generalized to polyelectrolyte brushes with more complicated
structures (e.g., chain architectures, heterogenous compositions and charge patterns) as well as brushes consisting of intrinsically disordered proteins. \cite{Zhulina_2010,McCarty_2019,Danielsen_2019,Yokokura_2023}

\section{Acknowledgements}

Acknowledgment is made to the donors of the American Chemical Society Petroleum Research Fund for partial support of this research. This research used the computational resources provided by the Kenneth S. Pitzer Center for Theoretical Chemistry.


\bibliography{Tex_Refs}

\end{document}